# Parametric Excitation of Internal Gravity Waves in Ocean and Atmosphere as Precursors of Strong Earthquakes and Tsunami[1]

**Sergey G. Chefranov[1], Alexander G. Chefranov[2]**

[1]A. M. Obukhov Institute of Atmospheric Physics RAS, Moscow, Russia
e-mail: schefranov@mail.ru
[2]Eastern Mediterranean University, Famagusta, North Cyprus
e-mail: Alexander.chefranov@emu.edu.tr
August 24, 2013

**Abstract.** The condition of internal gravity waves (IGW) parametric excitation in the rotating fluid layer heated from above, with the layer vibration along the vertical axis or with periodic modulation in time of the vertical temperature distribution, is obtained. We show the dual role of the molecular dissipative effects that may lead not only to the wave oscillations damping, but also to emergence of hydrodynamic dissipative instability (DI) in some frequency band of IGW. This DI also may take place for the localized in horizontal plane tornado-like disturbances, horizontal scale of which does not exceed the character vertical scale for the fluid layer of the finite depth. Investigated parametric resonance mechanism of IGW generation in ocean and atmosphere during and before earthquakes allows monitoring of such waves (with double period with respect to the period of vibration or temperature gradient modulation) as precursors of these devastating phenomena.

## *Introduction*

Development of high definition satellite radio-navigation systems allowing global monitoring of self shifts of the Earth surface elements leads to their active use for recording of evolution of coherent disturbances of the Earth surface related with precursors of and earthquakes themselves [1-5]. For better solving with these systems the problem of the strong earthquakes and tsunami forecasting, understanding of possibility of ionosphere responses on earthquakes and their precursors may play an important role [6-10]. In particular, in [10], a mechanism of lithosphere-ionosphere interactions due to the energy transfer by IGW generated in the atmosphere in the process of tectonic activity amplifying is discussed. Generating of IGW is possible due to modulation with time of the influx in troposphere of lithosphere gases and also due to related with it periodic variability of near-ground temperature gradient (or vibration by vertical of the respective ocean or atmosphere layer) providing possibility of IGW parametric resonance generation with double period [9, 10]. In particular, in [9], it is predicted possibility of parametric IGW generation with diurnal and two-day periods confirmed by later observations in various regions of the ocean [11, 12].

In the present work, generalization of theory [9] is made for the case of accounting of the fluid rotation as a whole that allows more particular consideration of the noted parametric mechanism of IGW generating in ocean and atmosphere on the rotating Earth in the f-plane approximation.

---

## Equations for small hydrodynamic disturbances in the Boussinesq approximation

1. Consider a flat unbounded with respect to horizontal rotating as a whole (with frequency $\Omega$) layer of viscous incompressible fluid of width H, having stable stratification along the vertical under constant values of temperature of the upper ($T_2$) and lower ($T_1$) layer boundaries (i.e. when $T_2 > T_1$). Let the layer in the result of activity of the external forces oscillates along the vertical as a whole with small amplitude $l$ for $\eta = \omega^2 l / g \ll 1$, where g is the gravitation acceleration.

Let's write equations for small hydrodynamic disturbances in Boussinesq approximation in the rotating with frequency $\Omega$ (for example, for the Earth rotation frequency $\Omega \approx 10^{-5} s^{-1}$) in the reference frame related with the lower layer boundary (axis z is directed upwards):

$$\frac{\partial u_r}{\partial t} - 2\Omega u_\varphi = -\frac{1}{\rho_0}\frac{\partial p'}{\partial r} + \nu(\Delta u_r - \frac{u_r}{r^2})$$

$$\frac{\partial u_\varphi}{\partial t} + 2\Omega u_r = \nu(\Delta u_\varphi - \frac{u_\varphi}{r^2})$$

$$\frac{\partial u_z}{\partial t} = -\frac{1}{\rho_0}\frac{\partial p'}{\partial z} + \nu \Delta u_z + \beta g T'(1 + \eta \sin \omega t) \quad (1)$$

$$\frac{\partial T'}{\partial t} + A u_z = \chi \Delta T'$$

$$\frac{\partial u_z}{\partial z} + \frac{1}{r}\frac{\partial (r u_r)}{\partial r} = 0$$

The system (1) is written in the cylindrical reference frame (z, r, $\varphi$) in axially symmetric case when velocity disturbances $u_r, u_\varphi, u_z$ (with respect to the rest), temperature $T'$ (with respect to linear profile of the temperature across the layer) and pressure $p'$ do not depend on the angular variable $\varphi$. In (1), the value of the equilibrium temperature gradient along the vertical, $A = (T_2 - T_1)/H$, and parameters $\nu, \chi, \beta$ are the coefficients of kinematic viscosity, temperature conductivity and thermal expansion respectively. The fluid density has the form $\rho = \rho_0(1 - \beta T')$, and temperature disturbances $T' = T - T_0(z), T_0 = T_1 + Az$, where T is the resulting temperature distribution in the fluid taking into account disturbances. Disturbed pressure is defined as $p' = p - \rho_0 \Omega^2 r^2 / 2$, where p – is the observed resulting pressure in the fluid.

We shall consider system (1) for simplicity in the case of free boundary conditions for z=0 and z=H (as it is shown in [9], in the case when the both boundaries are firm or one of the surfaces is firm, conclusions are qualitatively the same as for more idealized case of the free boundaries):

$$u_z = \partial^2 u_z / \partial^2 z = T' = 0; z = 0, H \quad (2)$$

2. Let us seek for the solution of the system (1) under the boundary conditions (2) in the following form:

$$u_r = \sum_{m=1}^{\infty} \sum_{n=1}^{\infty} B_{1nm}(t)\cos(\pi mz/H)Z_1(rk_n),$$

$$u_\varphi = \sum_{m=1}^{\infty} \sum_{n=1}^{\infty} B_{2nm}(t)\cos(\pi mz/H)Z_1(rk_n),$$

$$u_z = \sum_{m=1}^{\infty} \sum_{n=1}^{\infty} A_{1nm}(t)\sin(\pi mz/H)Z_0(rk_n), \quad (3)$$

$$p' = \sum_{m=1}^{\infty} \sum_{n=1}^{\infty} A_{2nm}(t)\cos(\pi mz/H)Z_0(rk_n),$$

$$T' = \sum_{m=1}^{\infty} \sum_{n=1}^{\infty} A_{3nm}(t)\sin(\pi mz/H)Z_0(rk_n),$$

where when considering horizontally localized disturbances in (3), functions describing dependence on the radial coordinate are defined via modified Bessel functions of zero and first order respectively: $Z_0 = K_0, Z_1 = K_1$. In the case of consideration of weakly damping for large r disturbances in (3), we have $Z_0 = J_0, Z_1 = J_1$, where $J_0, J_1$ are the usual Bessel functions of the zero and first order respectively.

## *Dissipative-centrifugal instability and tornado-like vortexes*

For the case of horizontally localized disturbances, after substitution of (3) in (1) (obviously, for (3), boundary conditions (2) always hold), we get the following system for the coefficients (further, we shall not show indexes n and m near these coefficients):

$$dB_1/dt - 2\Omega B_2 = k_n A_2/\rho_0 - \nu(\pi^2 m^2/H^2 - k_n^2)B_1$$
$$dB_2/dt + 2\Omega B_1 = -\nu(\pi^2 m^2/H^2 - k_n^2)B_2$$
$$dA_1/dt = \pi m A_2/\rho_0 H - \nu(\pi^2 m^2/H^2 - k_n^2)A_1 + \beta g A_3(1+\eta\sin\omega t) \quad . \quad (4)$$
$$dA_3/dt + AA_1 = -\chi(\pi^2 m^2/H^2 - k_n^2)A_3$$
$$\pi m A_1/H + k_n B_1 = 0$$

For weakly damping in the limit of large r disturbances in (4), it is necessary only to modify the sign near value $k_n^2$ (in the parenthesis near the coefficients of kinematic viscosity and temperature conductivity).

System (1) and its solution in the form of (3), (4) were considered earlier in [13] for the case of $\eta = 0$, i.e. when the layer vibration along the vertical axis is absent (or there is no periodic modulation in time of the linear temperature profile along the layer width). For $\eta = 0, \Pr = \nu/\chi = 1$, system (4), actually allows an exact solution in which all five unknown functions in (4) have the same dependence on time in the form:

$$T(t) = \exp(i\omega_1 t - 2\alpha_n t); A_1 = \operatorname{Re} A_{10}T, A_2 = \operatorname{Re} A_{20}T, A_3 = \operatorname{Re} A_{30}T, B_1 = \operatorname{Re} B_{10}T, B_2 = \operatorname{Re} B_{20}T;$$

$$B_{10} = i\omega_1 \pi m A_{30} / k_n HA, B_{20} = -2\Omega \pi m A_{30} / k_n HA, A_{10} = -i\omega_1 A_{30} / A, A_{20} = \frac{\rho_0 H(\omega_1^2 - N^2)A_{30}}{\pi m A};$$

$$N^2 = \beta g A; k_n^2 = \pi^2 m^2 (4\Omega^2 - \omega_1^2) / H^2(\omega_1^2 - N^2);$$

$$2\alpha_n = \nu(\frac{\pi^2 m^2}{H^2} - k_n^2) = -\frac{\nu \pi^2 m^2 (2\omega_1^2 - N^2 - 4\Omega^2)}{H^2(N^2 - \omega_1^2)}$$

(5)

where Re denotes operation of extraction of the real part, $N = \sqrt{-\frac{g}{\rho_0}\frac{\partial \rho(z)}{\partial z}}$ is the Brunt-Vaisala frequency for A>0 [9], and constant value $A_{30}$ is defined from the initial conditions on amplitude of the temperature field disturbances.

Thus, according to (5), it is possible to realize dissipative-centrifugal instability (DCI) caused by the presence of the finite molecular viscosity and sufficiently fast rotation of the fluid layer as a whole with threshold frequency

$$\Omega > \Omega_{th} = N/2 \tag{6}$$

and if values of disturbances' frequencies in (5) meet the following conditions

$$\sqrt{2\Omega^2 + \frac{N^2}{2}} > \omega_1 > N \; ; \tag{7}$$

then inline with negativeness of the value $2\alpha_n < 0$ in (5), also condition $k_n^2 > 0$ holds, Similar to (6) condition of DCI for two-dimensional oscillator in the rotating reference frame was obtained in [14] to explain observed cyclone-anti-cyclone vortex symmetry in the atmospheres of quickly rotating planets (Earth, Jupiter). In [15, 16], DCI representation suggested in [14], was used for investigation of large-scale motions in ionosphere and, in [17], it was used in relation with consideration of a linear instability mechanism leading to emergence of intensive tropic cyclones. Conditions (6), (7) provide exponential with time growth of tornado-like vortex disturbances having non-zero helicity.

### *Dissipative instability and IGW*

It is obvious that negativeness of $\alpha_n$ in (5) leading to the exponential growth of disturbances with time is possible also for relatively low-frequency disturbances corresponding to IGW for which the frequency does not exceed the value of the Brunt-Vaisala frequency. Then instead of conditions (6), (7), from (5), we get the following conditions of IGW generating:

$$N > \omega_1 > \sqrt{2\Omega^2 + N^2/2}, N > 2\Omega \tag{8}$$

Thus, presence of a stable stratification may lead to the effects similar to rotation of the system as a whole when realizing dissipative instability of localized over horizontal wave disturbances. Actually, the condition (8) even for $\Omega = 0$ provides exponential growth with time of the disturbances according to (5). But contrary to the condition of realization of DCI in [14], in (8), the value of the self frequency of the disturbance is bounded not only from above, but from below also.

## Parametric resonance mechanism of IGW generating

1. Let us consider the system (4) for non-zero modulation amplitude value $\eta$. Let for simplicity, again Prandtl number is equal to one (that is character for air, not for water, medium): $Pr = \nu/\chi = 1$. All unknown functions in (4) may be represented in the following form:

$$A_i = A'_i(t)\exp(-2\alpha_n t), i=1,2,3; B_k = B'_k(t)\exp(-2\alpha_n t), k=1,2; 2\alpha_n = \nu(\frac{\pi^2 m^2}{H^2} - k_n^2), \quad (9)$$

where contrary to (5), value $k_n^2$ has not a definite expression and may be positive (for the case of localized over horizontal disturbances), and also negative (for weakly damping disturbances in the limit of large r). In the latter case, it is necessary only to replace the sign before $k_n^2$ in (9).

From (4) and (9), we get for the amplitude $A'_3(t)$ (other amplitudes are expressed via it elementary):

$$d^2 A'_3/dt^2 + \omega_0^2(1+\varepsilon\sin\omega t)A'_3 = 0;$$
$$\omega_0^2 = \frac{N^2 + 4\Omega^2\pi^2 m^2/H^2 k_n^2}{1+\pi^2 m^2/H^2 k_n^2}, \varepsilon = \frac{\eta}{1+4\Omega^2\pi^2 m^2/N^2 H^2 k_n^2} \quad (10)$$

In (10), all the expressions are already the same for the cases of localized and weakly damping disturbances.

2. Using as in [9], Krylov-Bogolyubov averaging method, we get from (10) for $\varepsilon \ll 1$ in the region of the main demultiplicative parametric resonance frequency region:

$$A'_3 = a(t)\sin(\frac{\omega}{2}t + \phi);$$
$$da/dt = \gamma a\cos 2\phi, d\phi/dt = \delta - \gamma\sin 2\phi, \quad (11)$$
$$\delta = |\omega_0 - \frac{\omega}{2}| \leq \varepsilon, \gamma = -\varepsilon\omega_0/4$$

$$A'_3 = a(t)\sin(\frac{\omega}{2}t + \phi);$$
$$da/dt = \gamma a\cos 2\phi, d\phi/dt = \delta - \gamma\sin 2\phi, \quad (11)$$
$$\delta = |\omega_0 - \frac{\omega}{2}| \leq \varepsilon, \gamma = -\varepsilon\omega_0/4$$

The system (11) has non-adiabatic invariant

$$I = a^2(\delta - \gamma\sin 2\phi), \quad (12)$$

on existence of it in the parametric resonance region attention was firstly brought to in [9]. Actually, it is known that in the parametric resonance region, invariance of the adiabatic invariant [20-22]

$$J = E(t)/\varpi(t) = 4\pi bb^*, b = (idA'_3/dt + \varpi(t)A'_3)/\sqrt{2\varpi(t)}$$
$$b = |b|\exp(i\Psi), \varpi^2 = \omega_0^2(1+\varepsilon\sin\omega t) \quad (13)$$

representing ratio of the oscillator energy to its depending on time frequency, is broken. Invariant (12), contrary to the adiabatic invariant (13), corresponds to the discrete symmetry with respect to rotation on angles multiples of 180 degrees (or mirror reflections) because the value (12) does not change only under the following substitution. $\phi \to \phi + \pi n, n = 1, 2, ...$ . At the same time, for the adiabatic invariant existing out of the parametric resonance region, continuous symmetry of the gauge type is character (i.e. it remains not changed for arbitrary change of the value of the angle variable $\Psi$), leaving not-changed its value for any angle rotation [18]. Thus, in the region of parametric resonance frequencies, it takes place lowering of the symmetry (that is character for the phase transitions of the second kind) in the system described by an oscillator with varying with time frequency.

When invariant (12) is zero, the variable $\phi = const$, the amplitude a(t) may grow exponentially with time under condition $\gamma > \delta$ i.e. $a(t) \approx \exp(t\sqrt{\gamma^2 - \delta^2})$. It is obvious that for the possibility of parametric instability and exponential growth with time in (9), it is necessary that the following inequality is true

$$\sqrt{\gamma^2 - \delta^2} > 2\alpha_n \qquad (14)$$

Thus, accounting of rotation does not change qualitatively (and for not too large scales over horizontal, also, quantitatively) conclusions obtained in [9] on possibility of realization of parametric IGW generation mechanism in ocean (on the depths greater than the near surface layer of the order of 100 m) and in the upper atmosphere (in wave conducting stratosphere layers from 17 to 37 km, and in thermosphere, higher than 80km) in the regions for which it is character presence of positive temperature gradients and real-valued Brunt-Vaisala frequency. In particular, obtained in [9] estimates were confirmed by the data of IGW observation in the ocean [11, 12].

**Ionosphere precursors of the earthquakes and parametric IGW generating**

According to interferometer observations of the temperature variations by emission of atomic oxygen 630 nm in F2-region of ionosphere, it is noted essential temperature growth (by 350 K) and synchronous with it decreasing of emission intensity happening several dozen minutes before a strong (with magnitude M=7.5) Iranian Rudbar earthquake 20 June, 1990 [10]. On the west and east azimuths, temperature disturbances and emission intensities did not exceed measuring accuracy contrary to the pointed out above large values of these disturbances on the north and south azimuths. Thus, by an optical method, it was recorded local temperature growth of the upper atmosphere in two regions on the height of 270 km on the distance of 1000 km from the centre of the earthquake. Rayleigh waves could not cause such heating since the maximal heating took place long before the earthquake time. Also, acoustic-gravitational waves appearing in the atmosphere after coming there of a shock wave generated by an earthquake can't explain the phenomenon of the pointed out heating. In [10], as such a mechanism, it is considered intensification, several days prior to an earthquake, of non-stationary impulsive influx in atmosphere of lithosphere gases. This can cause generation in atmosphere of IGW with periods of about 1-2 hours and horizontal scales of several hundred kilometers. Such wave packets under favorable wind stratification may without essential energy losses penetrate with amplitude increasing to the heights of about 125 km. Higher, the waves substantially dissipate but also on the height of 200 km respective plasma disturbances related with Rayleigh-Tailor instability can cause forming of plasma bubbles (regions with lowered density) spreading to the heights of about 300 km and stretched by the azimuth along magnetic field lines. In such bubbles, essential polarized electric field emerges, effectively heating ions and also, as a consequence, the neutral atmosphere component [10]. According to estimates [19], agreeing with the observed data, the time of development of the plasma bubble is just less than 1 hour and the temperature of the neutrals increases by 300K.

In the considered in [19, 10] multistage lithosphere-ionosphere interaction, an important energy effective role might play IGW excitation parametric resonance mechanism considered in [9] and in the present work. Actually, the mechanism is effective not only for the first time IGW generation due to the periodic modulation of the near surface temperature gradient when lithosphere gases go out to the atmosphere but also for the second time IGW generation in vibrating along the vertical axis layer on the height of about 125 km where the first time generated IGW gets the maximal amplitude. Due to such second time parametric IGW generation, the IGW energy can most effectively be transmitted from the layer on the height of about 125 km to the layer of the height of about 200 km where development of the plasma instability starts. For example (see also [10]), for a layer of the width of 30 km, vertically oscillating on the height of 125 km with amplitude of 3 km, IGW with periods of about one to tens minutes are possible. In that relation, it is interesting to conduct spectral analysis of the disturbances spreading between the atmosphere layers on the heights of 125 and 200 km.